\documentclass[chicago,numberedappendix,tighten]{emulateapj}

\usepackage{amsmath,amssymb,natbib,graphicx}
\usepackage{color}
\usepackage{ulem}

\slugcomment{DRAFT \today}

\shorttitle{X-Ray emission from SN 2004dj}
\shortauthors{Chakraborti et al.}

\begin{document}


\title{X-Ray emission from SN 2004dj: A Tale of Two Shocks}		


\author{Sayan Chakraborti, Naveen Yadav, Alak Ray}
\affil{Tata Institute of Fundamental Research, 1 Homi Bhabha Road, Colaba, Mumbai 400 005, India}

\author{Randall Smith}
\affil{Harvard-Smithsonian Center for Astrophysics, 60 Garden Street, Cambridge, MA 02138, USA}

%\and

\author{Poonam Chandra}
\affil{Department of Physics, Royal Military College of Canada, Kingston, ON, K7K 7B4, Canada}

%\author{Our Friends}
%\affil{Other Institutes}

\email{sayan@tifr.res.in}


%\altaffiltext{1}{Institute for Theory and Computation,
%Harvard-Smithsonian Center for Astrophysics, 60 Garden St.,
%Cambridge, MA 02138, USA}


\begin{abstract}
Type IIP (Plateau) Supernovae are the most commonly observed variety of core collapse events.
They have been detected in a wide range of wavelengths from radio, through optical to
X-rays. The standard picture of a type IIP supernova has the blastwave interacting
with the progenitor's circumstellar matter to
produce a hot region bounded by a forward and a reverse shock. This region is thought to be
responsible for most of the X-ray and radio emission from these objects. Yet the origin
of X-rays from these supernovae is not well understood quantitatively. The relative contributions of particle
acceleration and magnetic field amplification in generating the X-ray and radio emission need
to be determined. In this work we analyze archival
Chandra observations of SN 2004dj, the nearest supernova since SN 1987A, along with published
radio and optical information. We determine the pre-explosion mass loss rate, blastwave velocity,
electron acceleration and magnetic field amplification efficiencies. We find that a greater
fraction of the thermal energy goes into
accelerating electrons than into amplifying magnetic fields. We conclude that the
X-ray emission arises out of a combination of inverse Compton scattering by non-thermal electrons
accelerated in the forward shock and thermal emission from supernova ejecta heated by the reverse
shock.
\end{abstract}





\section{Introduction}
Core collapse supernovae with hydrogen lines near maximum light
and a pronounced plateau in their visible band light curve that remain
within $\sim 1$ mag of maximum brightness for an extended period are
called type IIP supernovae.
The plateau duration is often 60-100 rest-frame days and is followed
by an exponential tail powered by radioactive decay at late times.
Type IIP supernovae constitute about 67\% of all core collapse supernovae
in a volume limited sample ($d< 30$ Mpc)
\citep{2009MNRAS.395.1409S}.
Their characteristic optical light curves are attributed to the hydrogen envelope of the progenitor
remaining largely intact before the core collapse.
Several lines of evidence suggest
that these stars were red supergiants when they exploded.
%These explosions are
%particularly suitable as probes of the
%formation of massive stars at high redshift
%(Cook et al 2009) and may be used as
%potentially good standardizable candles
%(alternate to type Ia supernova based
%cosmology,
%Poznanski et al, 2010)
%%; see however, D'Andrea
%%et al. 2010, ApJ 708, 661)
%which
%can refine measurements of the Dark Energy density and its
%associated equation of state parameter.
Type II supernovae (as well as the type Ib/c supernovae) are associated with
recent star forming regions of the spiral galaxies \citep{1997ARA&A..35..309F}
which suggest that their
progenitor stars are relatively short lived and therefore explode from
massive stars ($ M > 8 M_{\odot}$) which have faster burning of their
nuclear fuels at generally higher central temperatures compared to
less massive stars.
%They arise from core collapse of massive stars which
%are able to burn their fuels under
%hydrostatic equilibrium in non-degenerate conditions.
Their plateau brightness and duration combined with their expansion
velocities suggest a pre-supernova radius of typically red supergiant
dimensions (e.g. $10^{2\sim3} \; R_{\odot}$).
While there are more massive red supergiants in the Local Group of galaxies,
no high-mass red supergiant progenitors above $17 M_{\odot}$
have been found for IIP supernovae in direct detection efforts in
volume-limited search for supernova progenitor stars \citep{2009MNRAS.395.1409S}.
This {\it red supergiant problem} has led to a viewpoint that these
more massive stars could
have core masses high enough to form black holes and supernovae from them
could be too faint to have been detected. At the same time a
minimum stellar mass for a SN  IIP to arise from is about
$8.5 \pm 1.0 \; M_{\odot}$, consistent with the upper limit to white dwarf
progenitor masses. A census of type IIP progenitors
and the partial loss of their hydrogen envelopes
could potentially be important
for neutron star vs black hole formation and their number distribution
in the galaxy.

In a type IIP supernova, the expanding
ejecta interacts with the slow wind of
the red supergiant
even though the star retained most of its hydrogen envelope intact.
This interaction generates a hot region, bounded by the forward and reverse shocks,
which may emit thermal X-rays. These shocks may also accelerate charged particles
and synchrotron radiation from relativistic electrons
can lead to radio emission that is strong enough to be detected at
extragalactic distance scales. The measurable radio and X-ray
properties give complementary
information about the regions shocked by the forward and reverse
shocks and can constrain the physical properties of the interaction region and the
progenitor star.
Because type IIP supernovae have a plateau phase of high optical luminosity,
energetic electrons near the forward shock find a dense photon
environment of seed photons which may be Compton boosted to X-ray energies.
The resultant Compton cooling can affect the population of the
electrons which also emit at radio wavelengths and 
show suppressed radio fluxes above a characteristic cooling break until the
end of the plateau phase where the optical photon
density undergoes a rapid decline \citep{2006ApJ...641.1029C}.

Using SN 2004dj as a prototype of its class we would like to address the questions:
\begin{enumerate}
 \item What is the origin of X-rays detected by the Chandra from young type IIP supernovae?
 \item How fast was the progenitor losing mass via winds in the final phase before explosion?
 \item How efficiently does the supernova accelerate cosmic ray electrons in its forward shock?
 \item What is the extent of turbulent magnetic field amplification in the post-shock material?
 \item Are synchrotron and inverse Compton losses important in explaining the radio lightcurves?
\end{enumerate}

SN 2004dj, discovered by K. Itagaki \citep[see][]{2004IAUC.8377....1N}
on 31.76 July 2004 (UT) in the spiral galaxy NGC 2403, was the closest
\citep{2004AJ....127.2031K} normal type IIP SN \citep{2004IAUC.8378....1P}
observed to date. It was discovered
$50\pm21$ days \citep{2006AJ....131.2245Z} after its progenitor's core collapse and
had a peak magnitude of 11.2 mag in the V-band \citep{2006AJ....131.2245Z}.
Analysis of its pre-explosion images indicated that
the initial main sequence mass of the progenitor of SN 2004dj was
about $15 \; M_{\odot}$ \citep{2004ApJ...615L.113M}; see however
\citet{2005ApJ...626L..89W} where the progenitor had a
main-sequence mass of $12 M_{\odot}$.
SN 2004dj occurred at a position coincident with Sandage star number 96
(hereafter S96, and resolved by HST to be a young star cluster) in NGC 2403.
%
%X-ray emission from a SN is determined by the nature of the star that
%exploded as well the medium in which it exploded.
To date, only seven supernovae of type IIP are known X-ray emitters.
Only four of them are known radio emitters: Supernovae 1999em, 2002hh, 2004dj and 2004et.
X-rays from circumstellar interaction has been studied in detail for type
IIP supernovae including
SN 1999em \citep{2002ApJ...572..932P},
%1999gi, 2002hh, 2004dj,
and 2004et \citep{2007MNRAS.381..280M}.
%2006bp and SN 1994W (25 Mpc, 13.3 mag,
%this was an X-ray emitter (even at 1180 days after explosion),
%but it is
%the last one
%also classified as a narrow line plateau).
Because of its close range, SN 2004dj was easily detectable in the X-ray and radio.
The Chandra X-ray Observatory had targeted it multiple times early
after its explosion. The earliest detection was reported by \citet{2004IAUC.8390....1P}.

The X-ray and radio luminosities are both sensitive to the mass loss
rate and the initial mass of the progenitor star. In this
paper we demonstrate that for SN 2004dj the determination of the
inverse Compton powerlaw spectrum from the forward shock, the
characteristics of the X-ray line emitting thermal plasma from the reverse shock
and the radio light curve peak, uniquely determines some properties of the
explosion and the progenitor star. These include the pre-supernova mass
loss rate $(\dot M)$, the fraction of post-shock energy density which goes into
relativistic electrons $(\epsilon_e)$ magnetic fields $(\epsilon_B)$.
The time dependent nature of the thermal and non-thermal X-ray fluxes
also bear signatures of the circumstellar density profile as the
forward shock encounters more and more circumstellar matter and the
ejecta profile as the reverse shock
ploughs into the expanding ejecta. We also see the correlation of the
inverse Compton flux with the optical light curve.

\begin{figure}
 \includegraphics[angle=0,width=\columnwidth]{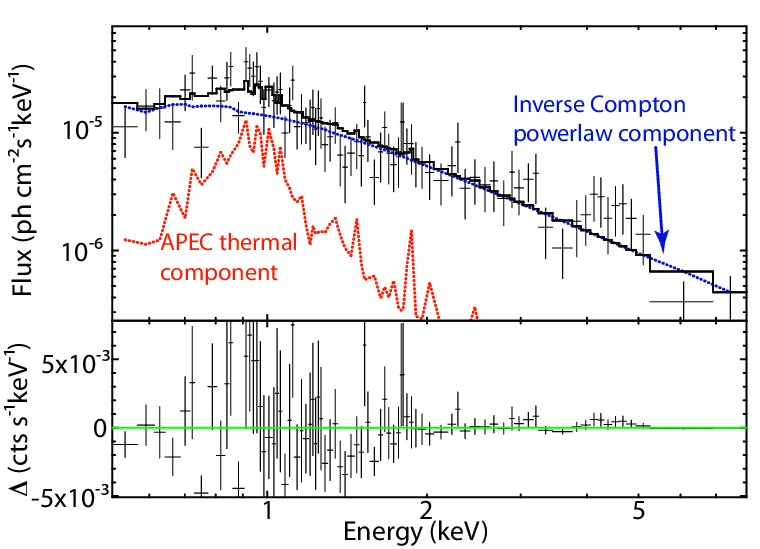}
 \caption{First X-ray spectrum of SN 2004dj on 2004 August 9. Bars are counts from Chandra,
 binned for display. Dotted line is the powerlaw model for the inverse Compton flux.
 Dashed line is the APEC model for the thermal plasma. Solid line is the full model.
 Note that the full model is dominated by the non-thermal flux even at $\sim1$ keV.}
 \label{spec1}
\end{figure}

\section{Observations of SN 2004dj}
As the nearest ($\sim3$ Mpc) supernova since SN 1987A, SN 2004dj enjoyed extensive
multi-wavelength coverage. The details of the observations in Radio, Optical
and X-rays, used in this work, are given below.

\subsection{Chandra X-Ray Observations}
SN 2004dj was observed by a Target of Opportunity program (PI: Walter Lewin,
Cycle: 5, ObsIDs: 4627-4630)
using the Chandra X-ray Observatory on four occasions:
2004 August 9 (See Fig \ref{spec1}), August 23, October 3 and December 22
(See Fig \ref{spec4}). The ACIS-S chips
were used on all occasions, without any grating, for 50 ks each. The supernova
was detected in the first of these observations
\citep{2004IAUC.8390....1P}. We analyze all four observations in this work
for the first time. See Table \ref{obstab} for details of observations.

\begin{table}[]
\centering
\caption{Observation summary of SN 2004dj with Chandra}
\begin{tabular}{l c r c}
\hline \hline
Date        & Exposure & Count Rate                  & Flux (0.5-8 keV) \\
(2004)      & (ks)      & ($10^{-3}$ sec$^{-1}$)      & (ergs cm$^{-2}$ s$^{-1}$)\\
\hline
Aug 09   & 40.9      & $12.80\pm 0.56$             & $8.81\times10^{-14}$ \\
Aug 23   & 46.5      & $10.03\pm 0.47$             & $6.98\times10^{-14}$ \\
Oct 03  & 44.5      & $5.60 \pm 0.36$             & $3.30\times10^{-14}$ \\
Dec 22 & 49.8      & $3.05 \pm 0.25$             & $2.02\times10^{-14}$ \\
\hline
\end{tabular}
\label{obstab}
\end{table}

Before spectral analysis, the data from separate epochs were processed separately
but in a similar fashion, following the prescribed threads from the Chandra Science
Center using CIAO 4.4 with CALDB 4.4.8.
The level 2 events were filtered in energy to reject all counts below 0.3 keV
and above 10 keV. The resulting events were mapped and the supernova easily identified.
The source region was marked off and a light curve was generated from the rest of the
counts. Flare times were identified in this light curve, as time-ranges where the count rate
rose above 3 times the rms. This resulted in a good time interval table which was used to
further filter out the events. The spectrum, response and background files were then
generated for these good events. In order to obtain the highest available spectral resolution,
the data were not binned; unbinned data were analyzed in the next steps.

\subsection{Radio Observations}
SN 2004dj has been observed extensively in radio bands from Aug 2004 until
May 2007 with various telescopes including the Very Large Array (VLA),
Giant Metrewave Radio Telescope (GMRT) and Multi-Element Radio Linked
Interferometer Network (MERLIN). MERLIN observed SN 2004dj extensively
in the 5 GHz band covering the period from 2004 August 5 to December 2. These
observations are reported in \citet{2005ApJ...623L..21B}. The GMRT observations
started on 2004 Aug 12 and
continued until 2007 May 22. The observations were made in the 1420,
610, 325 and 235 MHz bands. The first two GMRT observations in
1420 MHz band are reported in \citet{2004IAUC.8397....3C}.
The VLA followed SN 2004dj extensively starting from Aug 1, 2004 until
May 28, 2007. The observations were made in all VLA bands starting
from 1.4 GHz band upto 44 GHz band. The first 8.4 GHz observation was reported
by \citet{2004IAUC.8379....1S}. \citet{2006ApJ...641.1029C} discussed
these published radio observations of SN 2004dj to interpret the
physical properties of SN 2004dj and its interaction with the
circumstellar matter. They established Free Free Absorption to be the radio
absorption mechanism based on the 
published radio observations. A comprehensive paper including all the radio 
observations and
their detailed interpretation is underway (Chandra et al., 2012,
under preparation.)

%\subsection{Optical Observations}

\section{A Tale of Two Shocks}
Supernova ejecta hits the pre-explosion wind at a velocity much larger than its
characteristic sound speed. This creates a strong forward shock moving
into the circumstellar matter arising from the mass loss from the progenitor
\citep{1982ApJ...258..790C}. In case of a type IIP supernova, the material behind
this shock is too hot ($\sim100$ keV) and tenuous to contribute significantly to
the X-ray flux in the Chandra bands ($0.3-10$ keV) through its thermal emission. However,
the radio emission from type IIP supernovae is said to arise from the non-thermal electrons
accelerated in this region \citep{2006ApJ...641.1029C}. Some optical photons from a
supernova may also be inverse Compton scattered off energetic electrons in this region
into the X-rays \citep{2003A&A...397.1011S,2004ApJ...605..823B}. Scattering
from a non-thermal population of electrons with a power law index $p$ generates
a spectral index of $(p-1)/2$ in the optically thin radio synchrotron. So the inverse Compton
spectrum may be modeled as a power law with a photon index of $(p+1)/2$.

In the frame of the ejecta however, the circumstellar matter
impinges supersonically on the ejecta. According to \citet{1974ApJ...188..335M} this drives
an inward (in mass coordinates) propagating wave which reheats the ejecta, which would have
otherwise been adiabatically cooled by its rapid expansion, to around $\sim1$ keV.
A supernova typically ejects more matter at lower velocity. As the reverse
shock moves into slower moving ejecta, its heats up more gas which can contribute
to the X-ray flux. The plasma in this region can be modeled as a collisionally-ionized
diffuse hot gas.

\begin{figure}
 \includegraphics[angle=0,width=\columnwidth]{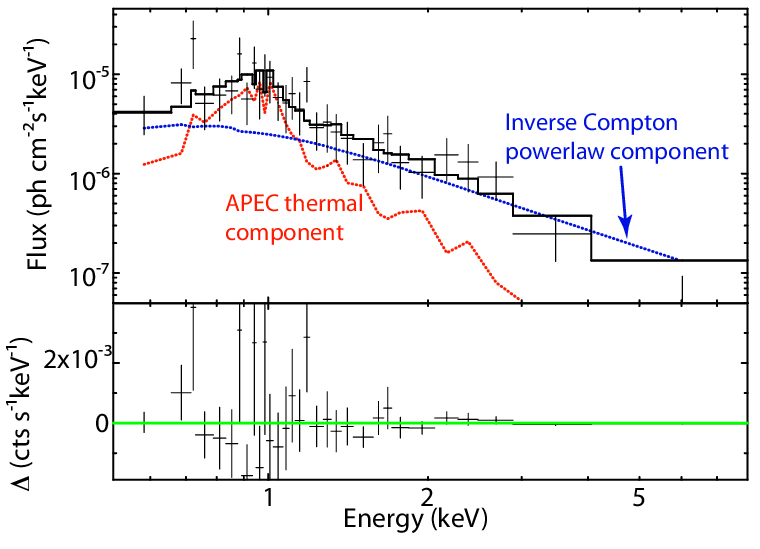}
 \caption{Last X-ray spectrum of SN 2004dj on 2004 December 22. Note that the relative
 contribution of the non-thermal inverse Compton, with respect to the flux from the
 thermal plasma has gone down, when compared with the first epoch. The full model
 is now dominated by the thermal flux at $\sim1$ keV.}
  \label{spec4}
\end{figure}

\subsection{X-ray Spectral Fitting}
We imported the extracted spectra into XSPEC 12.7.1 for further analysis. The data from all epochs
were jointly fitted with
the combination of a simple photon power law using {\it powerlaw} and collisionally-ionized
diffuse gas using {\it APEC} models \citep{2001ApJ...556L..91S},
passed through a photo-electric absorbing column using
{\it wabs}. The column density for wabs, plasma temperature for APEC and photon index for
powerlaw were shared between all epochs. The emission measure for APEC and normalization
for powerlaw were determined separately for each epoch from the joint analysis. As, the
fits were performed on unbinned data, each bin would have too few photons for the $\chi^2$
statistic to be useful. We therefore used the cstat statistic based on \citet{1979ApJ...228..939C}.
In order to evaluate the goodness of fit, we simulated 10000 spectra based on the best fit model
and found out that only 25\% of these simulations had cstat less than that for the data. We
therefore conclude that this model provides a good fit to the data. The best
fit absorption column density is
$n_{\rm H}=(1.7\pm0.5)\times 10^{21} \; {\rm cm^{-2}}$.
See Table \ref{obstab} for model fluxes on each date.

\subsection{Reverse-Shocked Material}
A self-similar solution for the interaction of the fast moving ejecta and the nearly
static (negligibly moving) circumstellar matter was found by \citet{1982ApJ...258..790C}.
This can be used to generate a complete description of the forward and reverse-shocked
material \citep{1985Ap&SS.112..225N}.

Following \citet{1982ApJ...258..790C} we consider supernova ejecta described by
\begin{equation}
 \rho_{\rm ej} \propto t^{-3} V^{-\eta},
\end{equation}
where the density falls due to free expansion and higher velocities have less matter, described
by the power law index $\eta$. This ejecta interacts with the circumstellar matter,
described by a power law profile
$\rho\propto r^{-s}$. For a steady wind ($s=2$), we have
\begin{equation}\label{rw}
 \rho_{\rm w} = \frac{A}{r^2} = \frac{\dot M}{4 \pi r^2 v_{\rm w}},
\end{equation}
where $\dot M$ is the mass loss rate and $v_{\rm w}$ is the wind velocity of the progenitor.
Here $A=\dot M / (4 \pi v_{\rm w})$ \citep{1982ApJ...258..790C}.
Shock jump conditions for a fast shock dictate that the density behind the forward shock is given by
$\rho_{\rm cs} = 4 \rho_{\rm w}$. In the thin shell approximation, this can be further
related to the density behind the reverse shock as
\begin{equation}\label{r0}
 \rho_0 = \frac{(\eta-3)(\eta-4)}{(3-s)(4-s)} \rho_{\rm cs}.
\end{equation}
According to \citet{1999ApJ...510..379M} $\eta$ is expected
to be 12 for a red supergiant progenitor. Using this value simplifies
the above expression to $\rho_0=36\rho_{\rm cs}=144\rho_{\rm w}$.

Therefore the reverse-shocked material is much denser than the circumstellar matter. So it
can cool efficiently via radiative processes and contribute to the thermal X-ray flux.

\begin{figure}
 \includegraphics[angle=0,width=\columnwidth]{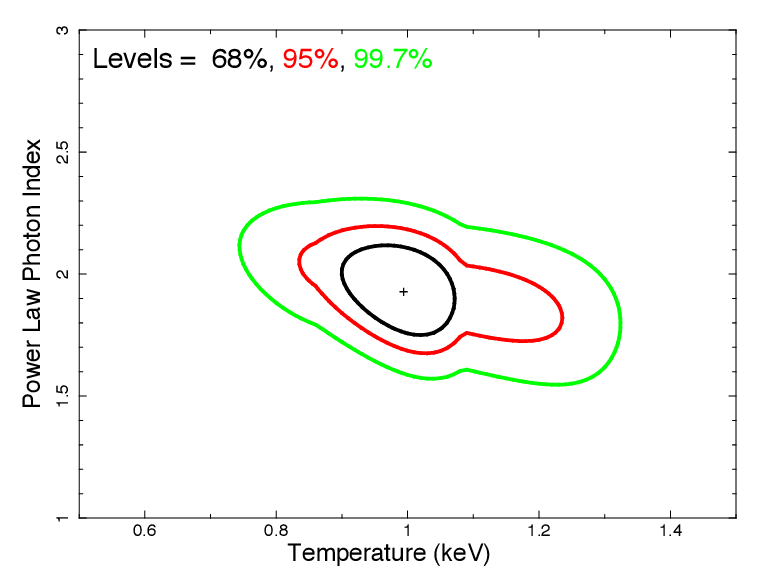}
 \caption{Confidence contours for Temperature (in the APEC model) vs Photon Index (in the
 powerlaw model). Note that the Chandra data can successfully break the degeneracy
 between the two components and account for their most important parameters,
 namely the temperature and photon index.}
 \label{T_ind}
\end{figure}

\subsection{Blastwave Velocity}
The velocity of a shock determines the energy imparted to particles crossing it, and hence
the post-shock plasma temperatures. Furthermore, the forward and reverse shock velocities are
connected by jump conditions once the ejecta and circumstellar matter profile are specified.
So, the post-shock temperature of ejecta which has crossed the reverse shock can be expressed
in terms of the blastwave velocity. Since the former is an observable, the latter can be
derived from it.

\citet{2006A&A...449..171N} give the temperature of the reverse-shocked material $T_0$, as
\begin{equation}
 T_0=2.27 \times 10^9 \mu \frac{(3-s)^2}{(\eta-s)^2} V_4^2 \; \;{\rm  K,}
\end{equation}
where $\mu$ is the mean atomic weight per particle,
$\eta$ is the power law index of the ejecta profile, $s$ is the power law index of the
circumstellar density profile and $V_4$ is the forward shock velocity in units of $10^4$ km
s$^{-1}$. $\mu$ is 0.61 in the solar-like outermost zone of the progenitor
\citep{2006A&A...449..171N}. $s$ is 2 for a circumstellar matter set up by
a steady wind, and we again use $\eta=12$.
With these assumptions, the forward shock velocity can be expressed as
\begin{equation}
 V_s = 10^4 \; \sqrt{\frac{kT_0}{1.19 \; {\rm keV}}} \; \; {\rm km s^{-1}}.
\end{equation}
Since, the best fit temperature is $0.997\pm0.054$ keV (See Fig \ref{T_ind}), the implied
velocity is $V_s=(9.2\pm0.3)\times 10^3$ km s$^{-1}$.

This is consistent with the fastest-moving ejecta in a typical type IIP supernova.
Therefore, the observed thermal component of the X-ray spectrum can be explained
as emission from reverse-shocked material, provided this is the blastwave velocity.
\citet{2007ApJ...662.1136C} have proposed diagnostics for circumstellar interaction
in Type IIP supernovae using the detection of high-velocity absorption features
in H$\alpha$ and He I lines during the photospheric stage. The highest velocity
H$\alpha$ absorption feature observed in the optical spectra of SN 2004dj is
at 8200 km s$^{-1}$
between 64 \citep{2007ApJ...662.1136C} and 102 days \citep{2006MNRAS.369.1780V}
with respect to the explosion date of June 28 proposed by \citet{2007ApJ...662.1136C}.
This overlaps with the epoch of Chandra observations and provides a
lower limit to the allowed forward shock velocity, so the velocity
that we obtain is consistent with constraints from optical spectroscopy.

\begin{figure}
 \includegraphics[angle=0,width=\columnwidth]{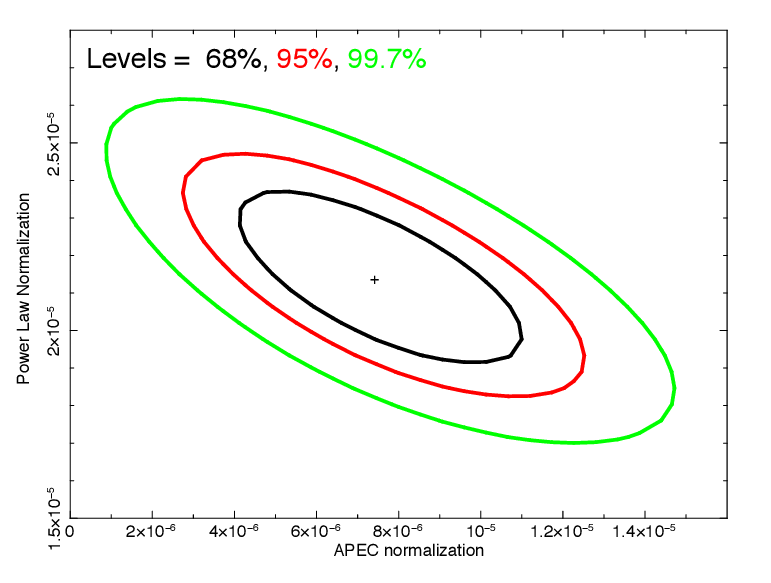}
 \caption{Confidence contours for normalization of the thermal flux (in the APEC model)
 the normalization of the non-thermal flux (in the powerlaw model). Note the
 anti-correlation between the thermal and non thermal fluxes, as their sum
 must explain the total flux which is fixed by observations.}
 \label{norms1}
\end{figure}

\subsection{Circumstellar Density}
The reverse-shocked matter contributes bulk of the thermal X-rays.
In order to find the circumstellar density it is necessary to estimate the amount of
reverse-shocked material in terms of the forward-shocked material.
%The density of the reverse
%shocked material has already been shown to be $144\rho_{\rm w}$, which gives us
Using Equations \ref{rw} and \ref{r0} and again assuming $\eta=12$, we get
\begin{equation}
 \rho_0 = \frac{36 \dot M}{\pi r^2 v_{\rm w}}.
\end{equation}
Similarly, following \citet{2006A&A...449..171N} we can express the total mass of
reverse-shocked material as
\begin{equation}
 M_0 = \frac{\eta-4}{4-s}M_{\rm cs}=\frac{4 \dot M R_s}{v_{\rm w}},
\end{equation}
where $R_s$ is the blastwave radius.
Multiplying these two results, we get,
\begin{equation}
 \rho_0 M_0 = \frac{144}{\pi} \left(\frac{\dot M}{v_{\rm w}}\right)^2\frac{1}{R_s}.
\end{equation}
These have to be converted to number densities to get the emission measure which
determines the thermal X-ray flux. For a solar-like composition appropriate for the
outermost shells of the supernova,
$\rho=1.17 {\rm amu} \times n_e = 1.40 {\rm amu} \times n_H$. Using these, we have
the emission measure for the reverse-shocked material as
\begin{equation}\label{em}
 \int n_e n_H dV = \frac{144}{\pi} \left(\frac{\dot M}{v_{\rm w}}\right)^2\frac{1}{(1.17 {\rm amu})(1.40 {\rm amu})R_s}
\end{equation}
We estimate the shock radius from the velocity determined in the previous Sub-Section as
$R_s=V_s t$, with the time calculated from the explosion date of 2004 June $11\pm21$
\citep{2006AJ....131.2245Z} or JD 2,453,168.
As all the other quantities are known, the circumstellar density can now be determined
from the emission measure.

The APEC fit to the thermal part of the X-ray spectrum of SN 2004dj gives,
\begin{equation}
 \frac{10^{-14}}{4 \pi [D_A (1+z)]^2}\int n_e n_H dv
 =(7.5\pm2.7)\times10^{-6}\;,
\end{equation}
for the first epoch (see Fig \ref{norms1}). NGC 2403 is nearby and has negligible redshift.
We adopt a distance to NGC 2403 of 3.06 Mpc based on Cepheids from \citet{2006ApJS..165..108S}.
Eliminating the emission measure between the last two equations,
we have,
\begin{equation}
 \frac{\dot M}{v_{\rm w}}=(2.0\pm0.7)\times10^{13}\; {\rm g\; s^{-1}}.
\end{equation}
Therefore the mass loss rate of the progenitor was
\begin{equation}
 \dot M = (3.2\pm1.1) \times 10^{-7} \left( \frac{v_{\rm w}}{10\; {\rm km/s}}\right) \; {\rm M_\odot \; yr^{-1}}
\end{equation}

This gives us $A$; however \citet{2006ApJ...651..381C} define a non-dimensional
$A_\star=A/(5\times10^{11}\; {\rm g\; cm^{-1}})$. For SN 2004dj, we therefore get
$A_\star=3.2\pm1.1$, which can now be compared with estimates for circumstellar density
from the radio emission.

%\begin{figure}
% \includegraphics[angle=270,width=\columnwidth]{T_nH.ps}
% \caption{Confidence contours for Temperature (in the APEC model) vs $n_H$ (in the
% wabs model).}
% \label{T_nH}
%\end{figure}

\subsection{Forward-Shocked material}
Electrons are believed to be accelerated to relativistic velocities in the
forward shock. These are in turn, responsible for the radio emission from supernovae.
The simplest model by \citet{1982ApJ...258..790C}, is to assume that a fraction
$\epsilon_e$ or $\epsilon_B$ of the thermal energy is used 
to accelerate electrons and amplify magnetic fields respectively. As a result
the radio emission from a supernova is dependent upon these fractions
and does not measure the circumstellar density directly. Instead, according to
\citet{2006ApJ...651..381C}, the radio emission measures
\begin{align}
 S_\star=& A_\star \epsilon_{B-1} \alpha^{8/19}=1.0 \left(\frac{f}{×0.5}\right)^{-8/19}
 \left(\frac{F_{\rm p}}{\rm mJy}\right)^{-4/19} \nonumber \\
 &\times\left(\frac{D}{\rm Mpc}\right)^{-8/19} \left(\frac{\nu}{5 {\rm \; GHz}}\right)^{-4/19}
 t_{10}^2,
\end{align}
where $F_{\rm p}$ is the peak flux at peak frequency $\nu$ at $10\times t_{10}$ days from
the explosion and $\epsilon_{B-1}=\epsilon_B/0.1$.
The equipartition factor is defined as $\alpha\equiv\epsilon_e/\epsilon_B$. 
Using the $4.99$ GHz, radio light curve from \citet{2005ApJ...623L..21B}
for SN 2004dj, we have $S_\star=5.1$.

The same relativistic electrons which generate the radio spectrum via synchrotron emission
may also contribute to the X-ray flux by inverse Compton scattering the optical photons
from the supernova's peak bolometric luminosity of around $\sim10^{42} {\rm \; ergs \; s^{-1}}$.
This may explain the dominant non-thermal component seen in the X-ray
spectrum at early times (see Fig \ref{spec1}). For an electron index $p=3$ this is expected
to generate a powerlaw with photon index 2, consistent with the observations
(see Fig \ref{T_ind}). Following \citet{2006ApJ...651..381C}
the normalization of the inverse Compton flux at 1 keV can be written as
\begin{align}
 E\frac{dL_{\rm IC}}{dE}&\approx8.8\times10^{36} \gamma_{\rm min} S_\star \alpha^{11/19} V_4 \nonumber \\
 &\times\left(\frac{L_{\rm bol}(t)}{10^{42} {\rm \; ergs \; s^{-1}}}\right)t_{10}^{-1} {\rm \; ergs \; s^{-1}},
\end{align}
where $\gamma_{\rm min}$ is the minimum Lorentz factor of the relativistic electrons.

The normalization of the inverse Compton flux obtained for the first epoch ($t_{10}\sim6$) of
Chandra observations, is seen to be $(3.8\pm0.5)\times10^{37}{\rm \; ergs \; s^{-1}}$
(see Fig \ref{norms1}).
Substituting this in the left hand side and using a $S_\star=5.1$
as found in this work, $V_4=0.92$ as implied by the temperature of the reverse-shocked plasma,
$L_{\rm bol}=0.89\times10^{42} {\rm \; ergs \; s^{-1}}$ from \citet{2006AJ....131.2245Z}, we
get
\begin{equation}
 \alpha\sim23\times \gamma_{\rm min}^{-19/11} ,
\end{equation}
where $\gamma_{\rm min}=1$ if the electron spectrum extends all the way down to those at rest.
However, if one only considers relativistic electrons with
$\gamma_{\rm min}\sim2.5$  as used by \citet{2006ApJ...651..381C} for SN 2002ap,
it implies $\alpha\sim4.8$. We prefer this latter (more realistic) assumption
as the bulk of the accelerated electrons cannot be at rest or very low velocities.
So we shall use this latter value in further calculations.

Now that we know $S_\star$ from radio synchrotron, $A_\star$ from thermal X-rays
and $\alpha$ from inverse Compton, we can use the definition of $S_\star$ to get
\begin{equation}
 \epsilon_B\equiv0.1\times\epsilon_{B-1}\sim0.082,
\end{equation}
which in turn implies
\begin{equation}
 \epsilon_e\equiv\alpha\times\epsilon_B\sim0.39 \; \;.
\end{equation}
Hence we have determined both $\epsilon_e$ and $\epsilon_B$ using a combination of
X-ray, radio and optical data.

\begin{figure}
 \includegraphics[angle=0,width=\columnwidth]{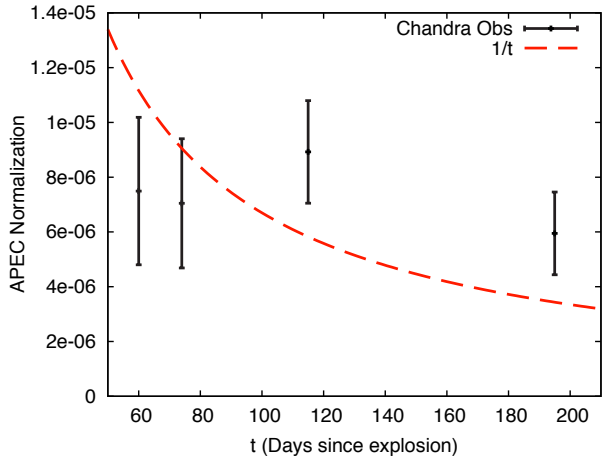}
 \caption{Time evolution of the normalization of the thermal APEC component (proportional to
 the emission measure) in the spectrum of SN 2004dj obtained from simultaneous fits to
 Chandra spectra. The time interval is measured with respect to the explosion date of June 11 from
 \citet{2006AJ....131.2245Z}. Line is best fit $\propto t^{-1}$ function. Note that the data
 do not rule out a constant mass loss rate model, but suggest one with variable
 mass loss.}
 \label{t_lc}
\end{figure}

\section{Time Evolution}
The first things that are apparent from the Chandra observations of SN 2004dj is that
the total X-ray flux softens and falls with time. These gross features
have already been observed in another type IIP supernova, SN 2004et by
\citet{2007ApJ...666.1108R}. However, the quality of the SN 2004dj spectra allows
a quantitative explanation for the first time. While both the thermal and non-thermal
components fall eventually, their proportion changes with time. At early times, the
spectra is dominated by the inverse Compton flux and is therefore harder (see Fig \ref{spec1}).
As the bolometric light curve of the supernova decays, the source of seed photons to be
scattered turns off. So at late times, the spectra is dominated by emission from
the reverse-shocked thermal plasma and is therefore softer (see Fig \ref{spec4}).

\subsection{Thermal X-Rays}
Since the post shock temperature in a self similar blastwave does not change much,
as the shock velocity is slowly varying, the variation
in the thermal X-ray flux would come from the variation in the emission measure. We have
already derived in Equation \ref{em} that the emission measure scales as $1/R_s$. Therefore,
if the radius scales approximately as $R_s=V_s t$, then we have
\begin{equation}
  \int n_e n_H dV \propto t^{-1},
\end{equation}
for a constant pre-explosion mass loss rate.
We comparing this to the temporal variation of the normalization of the APEC flux from SN 2004dj
(see Fig \ref{t_lc}). While a constant mass loss rate cannot be ruled out, there is a hint
of variation up to 50\% from a constant mass loss.

\subsection{Inverse Compton}
\begin{figure}
 \includegraphics[angle=0,width=\columnwidth]{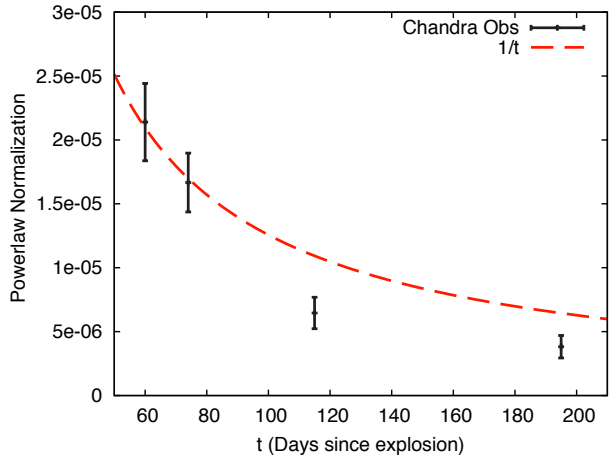}
 \caption{Time evolution of the normalization of the non-thermal inverse Compton component
 in the spectrum of SN 2004dj obtained from simultaneous fits to
 Chandra spectra. Line is best fit $\propto t^{-1}$ function. The time interval is measured
 similarly as in Fig \ref{t_lc}. Note that the first
 two epochs, for which $L_{\rm bol}$ is nearly constant, lie on the line. Also note
 that one the $L_{\rm bol}$ decreases due to the termination of the plateau phase
 beyond 100 days, the inverse Compton flux also falls off.}
 \label{ic_lc}
\end{figure}

The time variation in the inverse Compton flux comes from the expansion if the blastwave and
the dimming of the supernova's supply of seed photon. 
\citet{2006ApJ...651..381C} have shown that the normalization of the inverse Compton flux scales as
\begin{equation}
 E \frac{dL_{\rm IC}}{dE}\propto\frac{L_{\rm bol}(t)}{t}.
\end{equation}
Thus the inverse Compton flux is expected to behave as $\propto t^{-1}$ in the plateau phase
where $L_{\rm bol}$ is nearly constant for a type IIP supernova. Thereafter, as the supernova
luminosity falls off, the inverse Compton flux is expected to fall off from this behavior.
This prediction is borne out by the observations as the first two epochs (see Fig \ref{ic_lc})
which lie in the plateau phase can be fitted by a $\propto t^{-1}$ function, while the late
time flux normalization falls off significantly below this curve.

\section{Implications}
Our results have wide ranging implications for the progenitors of type IIP supernovae,
conditions of the plasma in the supernova shock waves, particle acceleration and
magnetic field amplification schemes. These are discussed briefly below.

\subsection{Progenitors of type IIP Supernovae}
According to \citet{2004ApJ...615L.113M} the progenitor of SN 2004dj had a mass of
$\sim15 {\rm M_\odot}$. In this mass range the progenitor dies as a red supergiant
and \citet{2006ApJ...641.1029C} find theoretical mass loss rates in the range
$\dot M= (0.84-1.6)\times10^{-6} \; {\rm M_\odot \; yr^{-1}}$ for such progenitors.
However, we find pre-explosion mass loss rate for the progenitor of SN 2004dj
to be
$\dot M = (0.32\pm0.11) \times 10^{-6} \left( {v_{\rm w}}/{10\; {\rm km \; s^{-1}}}\right) \; {\rm M_\odot \; yr^{-1}}$.
This can be reconciled if the wind velocity is $\gtrsim26\; {\rm km \; s^{-1}}$, which
is unlikely as the wind velocity of a typical red supergiant is around $10-15\; {\rm km \; s^{-1}}$.
However, these rates have been calculated for stars of solar metallicity.
\citet{2009MNRAS.395.1409S} suggest that the metallicity of the SN 2004dj site
is ${\rm log[O/H]}=8.4$. Assuming a solar value of 8.7, this implies a host metallicity
of $Z=0.5Z_\odot$. Using $\dot M \propto Z^{0.5}$ following \citet{1992A&AS...96..269S,2003ApJ...591..288H}
would bring down the theoretical values closer to the observed one.

Stars in the mass-range 11-19 $M_{\odot}$ were evolved using
MESA \citep{2011ApJS..192....3P} for three different metallicities
z=0.4, 0.50 and 0.75 $Z_\odot$. The mass loss scheme that was
used is described as the {\it Dutch} Scheme in the MESA code; see Sec 6.6
MESA code paper by \citet{2011ApJS..192....3P}.
This scheme switches on a RGB wind depending upon
the burning stage. The scheme takes care of the surface temperature
changes in the different stages of stellar evolution and adjusts itself
accordingly. The mass loss in the RGB branch, which is relevant
for the final stages of type IIP progenitors, follows
\citet{1988A&AS...72..259D}.
The stars were evolved till the central density had reached $10^{12}$
g cm$^{-3}$. The calculated mass loss was the average value over
the last century of evolution. This is the mass which the blastwave
sweeps up during first few years of its evolution.

\begin{figure}
 \includegraphics[angle=0,width=\columnwidth]{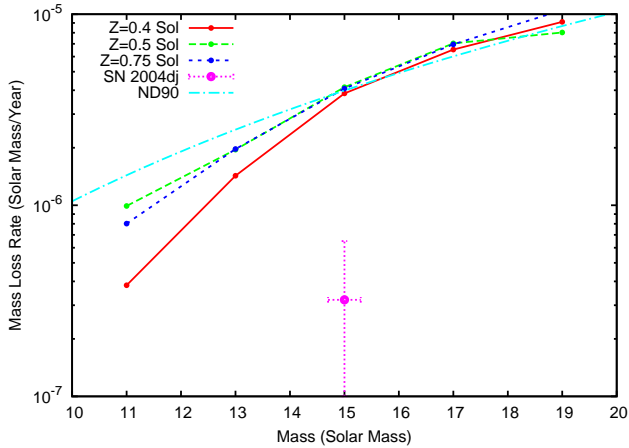}
 \caption{Zero Age Main Sequence Mass (ZAMS) and Mass Loss rate (in the last
 decades of their lives) for the progenitor of SN 2004dj (Magenta Cross), MESA runs of 0.4
 (solid red), 0.5 (dotted blue) and 0.75 (dashed green) $Z_\odot$ metallicity, and theoretical
 line (dot-dashed cyan) from \citet{1990A&A...231..134N} (with $R=10^3R_\odot$) plotted
 for comparison. Error bars are $3\sigma$ confidence intervals. Note the disagreement
 between the observed properties of the SN 2004dj progenitor and the theoretical models.}
 \label{mmdot}
\end{figure}

In Fig \ref{mmdot} we compare the progenitor mass of SN 2004dj as determined by
\citet{2004ApJ...615L.113M} and mass loss rate determined in this work, with the
expectation from stellar evolution and standard mass loss rate prescriptions.
There is an obvious disagreement between the observed mass loss rate and theoretical
predictions. Such a disagreement was also noted by \citet{2007ApJ...662.1136C} who
determined mass loss rates for the SN 1999em and SN 2004dj using absorption lines
from optical spectroscopy.

\subsection{Ionization Equilibrium}
The analysis of X-ray spectra to yield parameters for the underlying
explosion models of young supernovae and SNRs is significantly complicated by
the possibility that their X-ray emitting plasmas may not be
ionization equilibrium. The X-ray emission comes from an impulsively
heated (by the blastwave or the reverse shock) uniform and homogeneous
gas which is initially cold and neutral. For young extragalactic supernovae,
which are point sources contained within the telescope beam-size, the
X-ray spectra are spatially averaged over the whole object. In the
case of non-equilibrium plasma the spectra depend upon the shock
velocity $V_s$, the ionization timescale $\tau_0$ (the product of the
supernova's age $t$ and the post-shock electron number density $n_{e}$)
and the extent of electron heating at the shock front (which determine
how different the electron and ion temperatures $T_e, T_s$ are).
Spectra consisting of lines whose strengths depend upon the heavy
elements abundances and their relative fractions. Both X-ray continuum
and lines arise from electrons undergoing bremsstrahlung or electron
impact excitation of ions. \citet{2001ApJ...548..820B} have
emphasized that for SNRs ionization equilibrium models are often not
adequate, and the constant-temperature, single-ionization timescale
NEI model is better than an equilibrium model. Young supernovae have shorter
lifespans till their present age but may be expanding through denser
circumstellar media and so these issues of adequacy of NEI vs IE
models may or may not be relevant.

%\begin{figure}
% \includegraphics[angle=270,width=\columnwidth]{Tau_T.ps}
% \caption{Confidence contours for $\tau_0$ (in the NEI model) vs the temperature.
% Note that while the temperature is constrained, the ionization timescale
% is not constrained at all. Therefore, the NEI model should not be used for SN 2004dj.}
% \label{Tau_T}
%\end{figure}

Because substantial X-ray data from Chandra exists in multiple epochs
for this target, we
investigate the relevant parameter space to see whether NEI models would
constrain emission model parameters sufficiently well to require their
use (as opposed to IE models).
The NEI model is an XSPEC provides a non-equilibrium
ionization collisional plasma model which uses APED \citep{2001ApJ...556L..91S}
to calculate the resulting spectra. It assumes a constant temperature
and single ionization timescale parameter. In NEI models the ionic
states of each abundant element (e.g., C, N, O, Ne, Mg, Si, S, Ca, Fe,
Ni) are solved for through time-dependent ionization equations.
%which
%are often solved efficiently by the use of the eigenfunction method
%(Masai 1984, Hughes and Helfand 1985, etc). These calculations (e.g.
%Borkowski et al 2001) include collisional ionizations by electrons,
%both direct and through excitations followed by auto-ionizations and
%radiative and dielectronic recombinations and start calculations with
%all elements except for H and He to be neutral initially.
APEC \citep{2001ApJ...556L..91S}, which is used throughout this work,
on the other hand does an
Ionization Equilibrium calculation for
X-ray emission spectra of
collisionally-ionized diffuse gas calculated using the ATOMDB code
v2.0.1 (http://www.atomdb.org/) that includes a set of relevant ions and
their lines and perform a population calculations.
The atomic data and line transitions information are at a
more comprehensive state than other  models like Mekal and has a
wavelength coverage from IR to X-ray.

However, when the resulting grid of feasible values for $\tau_0$
is sampled
%(see Fig \ref{Tau_T}),
it was noted that the data does not constrain it at all.
In case of SN 2004dj, the post reverse shock density at the first
observation epoch is $\sim5\times10^6 \; {\rm cm^{-3}}$, and the age
is $5\times10^6$	s. This will give a $\tau_0> 10^{13} \; {\rm s \; cm^{-3}}$, which 
should be enough time for the plasma to come to ionization equilibrium.
This may not be strictly true for the material behind the forward shock,
however it would be too hot and too low density to contribute significantly
to the Chandra spectrum. Therefore it is safe to make the assumption of
ionization equilibrium while modeling the Chandra X-ray spectrum of SN 2004dj,
although higher resolution data such as will be available with Astro-H
might reveal NEI features.

\subsection{Particles and Magnetic Fields}
Radio or X-ray light curves of supernovae are often explained using varying
amounts of accelerated electrons and amplified magnetic fields. However it is
difficult to ascertain the relative contribution of each. By modeling the
observed X-ray spectrum as a combination of thermal and inverse Compton flux,
we have been able to determine the equipartition factor for SN 2004dj. Our
result shows that the plasma is away from equipartition, with electrons having
more of the total energy than magnetic fields. Since $\epsilon_e$ and $\epsilon_B$
have also been found individually, theories of particle acceleration and 
magnetic field amplification can now be tested against them. This is an
unique opportunity.

Since the electrons and magnetic fields are responsible for producing the
synchrotron spectra of supernovae, it is important to look breaks in the
radio spectrum of a young supernova. \citet{2004ApJ...604L..97C} associate
such a break in SN 1993J (not a type IIP, but a type IIB supernova) with
the phenomenon of synchrotron aging of radiating electrons. The magnetic
field in the shocked region can be calculated from the break in the spectrum,
independent of the equipartition assumption between energy density of
relativistic particles and magnetic energy density. The synchrotron cooling
break appears at a certain frequency when electrons radiating above that
frequency have lost a significant portion of their energy to synchrotron
emission. Demanding that the synchrotron loss time scale is equal to
the age of the supernova, we can use Equation 5 of \citet{2006ApJ...641.1029C}
to express the frequency of the synchrotron cooling break as
\begin{align}
 \nu_{\rm Syn} = & 240 \; \left(\frac{\epsilon_B}{0.1}\right)^{-3/2} 
  \left(\frac{\dot M}{10^{-6}{\rm \; M_\odot \; yr^{-1}}}\right)^{-3/2} \nonumber \\
  &\times\left(\frac{v_{\rm w}}{10\; {\rm km \; s^{-1}}}\right)^{3/2}
  \left(\frac{t}{60\; {\rm days}}\right)
  {\rm \; GHz,}
\end{align}
which has the same temporal evolution as derived in \citet{2011ApJ...729...57C}.
For the epochs at which it was observed in the radio
This is much higher than the frequencies accessible with the EVLA, therefore
synchrotron cooling will be unimportant for studying the radio light curves
of SN 2004dj.
However for future studies of young type IIP supernovae, these frequencies may
become detectable with mm wave observatories such as the SMA or ALMA.

Another important mechanism for electron cooling is inverse Compton losses
against the low energy photons from the supernova photosphere. Due to
this process high energy electrons may lose bulk of their energy in boosting
the seed photons to X-rays, thereby contributing to the X-ray flux, as discussed
in this paper for SN 2004dj. Demanding that the inverse Compton loss time
scale is equal to the age of the supernova, we can use Equation 6 of
\citet{2006ApJ...641.1029C} to express the frequency of the inverse Compton
cooling break as
\begin{align}
 \nu_{\rm IC} = & 8 \; \left(\frac{\epsilon_B}{0.1}\right)^{1/2} 
  \left(\frac{\dot M}{10^{-6}{\rm \; M_\odot \; yr^{-1}}}\right)^{1/2} \nonumber \\
  &\times\left(\frac{v_{\rm w}}{10\; {\rm km \; s^{-1}}}\right)^{-1/2}
  \left(\frac{t}{60\; {\rm days}}\right) \nonumber \\
  &\times\left(\frac{V_s}{10^4\; {\rm km \; s^{-1}}}\right)^{4}
\left(\frac{L_{\rm bol}(t)}{10^{42} {\rm \; ergs \; s^{-1}}}\right)
  {\rm \; GHz.}
\end{align}
For the determined relevant values of the parameters determined in this work,
the $\nu_{\rm IC}$ is within the frequency coverage of the VLA. Therefore,
any interpretation of the radio light curve of SN 2004dj should account
for inverse Compton losses and expect to see such a cooling break.

\section{Conclusions}
In this work we have used SN 2004dj as a prototype for type IIP supernovae
to answer some fundamental questions about their circumstellar interaction.
\begin{enumerate}
 \item What is the origin of X-rays detected by the Chandra from young type IIP supernovae?
 The X-rays detected by Chandra arise from a combination of thermal and non-thermal processes.
 They are dominated at early times by optical supernova photons
 inverse Compton scattered by relativistic electrons at the forward shock which
 falls off as $\propto L_{\rm bol}/t$. At late times
 it is dominated by thermal emission from reverse shock heated plasma which only
 falls as $\propto t^{-1}$.
 \item How fast was the progenitor losing mass via winds in the final phase before explosion?
 The progenitor was losing mass at a rate of
 $\dot M = (0.32\pm0.11) \times 10^{-6} \left( {v_{\rm w}}/{10\; {\rm km \; s^{-1}}}\right) \; {\rm M_\odot \; yr^{-1}}$.
 This is less than the value expected from mass loss prescriptions in the literature
 for the putative $15 {\rm M_\odot}$ progenitor.
 \item How efficiently does the supernova accelerate cosmic ray electrons in its forward shock?
 Around a third ($\epsilon_e\sim0.39$) of the energy thermalized by the collision of the ejecta with the
 circumstellar matter is used in accelerating electrons to relativistic velocities.
 \item What is the extent of turbulent magnetic field amplification in the post-shock material?
 Around a tenth ($\epsilon_B\sim0.082$) of the thermal energy available is used in the turbulent amplification of
 magnetic fields.
 \item Are synchrotron and inverse Compton losses important in explaining the
 radio lightcurves? In the case of SN 2004dj synchrotron losses are unimportant
 while inverse Compton losses will have to be taken into account while explaining
 its radio lightcurve.
\end{enumerate}

\acknowledgments
This research has made use of data obtained from the Chandra Data Archive and software provided
by the Chandra X-ray Center (CXC) in the application packages CIAO and ChIPS.
%We thank Walter Lewin for proposing these observations and correspondence
%and advice prior to the submission of this paper.

%Not decided.

% PUT FIGURES HERE

%\clearpage

%\clearpage

%\clearpage




%\clearpage

%\bibliographystyle{apj}
%\bibliography{04dj}

\end{document}